\documentclass[aps,prb,twocolumn,showkeys]{revtex4-1}
\usepackage[tbtags]{amsmath}
\usepackage{tikz-feynman}
\usepackage{graphics,graphicx,epsfig,amssymb,xcolor,hyperref,latexsym,
float,bm,bbm,times,adjustbox,verbatim,subfigure}
\newcommand{\at}[2][]{#1|_{#2}}

\begin{document}

\title{Spatial behavior in a Mott insulator near the voltage 
driven resistive transition}

\author{Arijit Dutta and Pinaki Majumdar}

\affiliation{Harish-Chandra Research Institute,
HBNI, Chhatnag Road, Jhunsi, Allahabad 211019, India}
\date{\today}


\begin{abstract}
We develop a real space theory of the voltage bias driven transition from a 
Mott insulator to a correlated metal. Within our Keldysh mean field approach 
the problem reduces to a self-consistency scheme for the charge and spin 
profiles in this open system. We solve this problem for a two dimensional 
antiferromagnetic Mott insulator at zero temperature. The charge and spin 
magnitude is uniform over the system at zero bias, but a bias $V$ leads to 
spatial modulation over a lengthscale $\xi(V)$ near the edges. $\xi(V)$ grows 
rapidly and becomes comparable to system size as $V$ increases towards a 
threshold scale $V_c$. The linear response conductance of the insulator is 
zero with the current being exponentially small for $V \ll V_c$. The current 
increases rapidly as $V \rightarrow V_c$. Beyond $V_c$, we observe an 
inhomogeneous low moment antiferromagnetic metal, and at even larger bias
a current saturated paramagnetic metal. We suggest an approximate scheme 
for the spectral features of this nonequilibrium system.
\end{abstract}


\maketitle

\section{Introduction}

The field driven breakdown of band insulators is a well understood 
phenomenon, owing to the early work of Zener \cite{zener1934}.
In these insulators electron correlation effects are neglected and
the breakdown is understood in terms of 
field assisted quantum tunneling of electrons across 
the band gap. In contrast, Mott insulators are strongly 
correlated, with a charge gap of many body origin 
due to
Coulomb repulsion between electrons. The nonequilibrium physics of Mott 
insulators has attracted a lot of attention in recent times 
with both experimental
\cite{wang,okamoto-nmat-2017,radu,nakamura2013,cario2013,
zimmers,kawasugi,toda,sabeth,wu,cario2008,kumai1999}
and theoretical
\cite{aoki-dmrg-2005,freericks2006,mazza2015,mazza2016,aoki-tdschr-2003,
dagotto2010,eckstein2010,aron-prl2012,okamoto2008,aron-prl2015,
fabrizio2010,han,li,rubtsov2016,tanaka} 
studies trying to explore 
the effect of the strong electron correlation 
on nonlinear transport in these insulators.

Experiments have probed Mott insulators out of equilibrium by either 
applying a dc bias\cite{kumai1999,nakamura2013,wang,zimmers,radu,wu}, 
by photoexcitation\cite{toda},
or using a pulse source\cite{okamoto-nmat-2017,sabeth,cario2008,cario2013}, 
and measured both ultrafast as well as long time (steady state) 
response.  In most of the experiments the current
voltage (I-V)  characteristic at low temperature 
has a sharp threshold voltage 
\cite{sabeth,kumai1999,zimmers,radu,nakamura2013,wu}.
The theory for band insulators 
predict that the I-V is given by the Landau-Zener
(LZ) form $I \sim V e^{-V_{th}/V}$, with $V_{th} \propto \Delta^{2}L/W$ where 
$\Delta$ is the band gap, $L$ is the longitudinal size and $W$ 
the bandwidth of the system. The observed I-V characteristics
at Mott breakdown are quite different from the LZ response,
and show a sharp threshold at low temperature.

The theoretical studies broadly use two approaches: 
field driven and bias driven. In the
field driven approach a constant electric field is applied across the
system, either by introducing a time dependent gauge field via Peierls
coupling or by imposing a linear potential gradient.
This has been used in the one dimensional Hubbard model 
to study the 
time-dependent Schrodinger equation \cite{aoki-tdschr-2003}, 
and also in an application of the 
density matrix renormalisation group (DMRG)\cite{aoki-dmrg-2005}.
Dynamical mean field theory (DMFT)\cite{georges96} generalised to
nonequilibrium situations\cite{freericks2006,eckstein2010,aron-prl2012,
aoki-ndmft-2014,aron-prl2015} has  extensively used this approach.
Many of these studies find that the I-V characteristics obey the
LZ form. Recently, the Hartree-Fock mean field approach
has also been used to study field driven problems. A discontiuous
insulator to metal transiton, along with a region of bistability,
has been found within a homogeneous mean field study\cite{han}.
For disordered systems, both hysterisis and filamentary conduction
have been found within an unrestricted Hartree-Fock study\cite{li}.

In the bias driven setup the interacting `bulk' 
is coupled to noninteracting
leads across the sample. A chemical potential difference 
(bias) between the two leads tends to drive a current.
Time dependent DMRG\cite{dagotto2010}, nonequilibrium DMFT
\cite{okamoto2008,mazza2016}, and a 3D time depnedent
Gutzwiller mean field based study \cite{fabrizio2010,mazza2015} 
have used this setup.
These methods {\it impose} a linear gradient across the system
or construct a screening potential by hand\cite{okamoto2008}.
Despite the assumed form for the potential the Gutzwiller method
obtains spatially inhomogeneous behaviour in various quantities.
The current in these studies do not show the sharp
voltage driven change that is observed in experiments
\cite{sabeth,kumai1999,zimmers,radu,nakamura2013,wu}, and fit 
rather with the LZ form.
A few bias driven calculations, however, do compute the
internal field self consistently and find that the
breakdown process is preceded by a spatially modulated
state \cite{rubtsov2016,tanaka}. This hints that the 
{\it spatial symmetry breaking} due to the applied bias
can promote inhomogeneous states which play a crucial
role in the breakdown.

To explore this aspect we used
a Keldysh mean field approach to study the
2D Hubbard model, at half-filling and strong interaction, 
connected to metallic leads. 
Our main finding is on the non trivial spatial 
behaviour of the charge and spin density in the 
Mott insulator as the system heads towards 
breakdown.  
We observe that there is a weakly size dependent 
threshold voltage $V_c$, of order the gap in the
zero bias Mott insulator, which defines the reference scale in
the biased problem. Around $V = V_c$ the system
shows crossver from exponentially small current 
to a high current state. 
The key features of this phenomenon can be captured
by a bias dependent `penetration length', $\xi(V)$,
which becomes comparable to system size as $V \rightarrow V_c$.

We uncover the following progression as the 
bias is increased across the Mott insulator.
In our notation
AF-I and AF-M are antiferromagnetic insulator and metal, 
respectively, and PM-M is paramagnetic metal.
%
\begin{enumerate}
\item
Pre-breakdown AF-I ($V \ll  V_c$):
In this window the charge and spin fields, $\phi_i$ 
and $M_i$, are weakly modified near the edge, from the $V=0$ 
value.  The deviation $\delta \phi_i$ and $\delta M_i$ 
decay as  $\sim e^{-R_i/\xi}$ into the bulk. There are 
exponential tails in the subgap density of states and a 
current $I \sim e^{-V_{c}/V}$.
\item 
Transition from AF-I to AF-M ($ V \sim V_c$):
The lengthscale $\xi(V)$ `diverges' as $V \rightarrow V_c$,
and the current rapidly rises over a small voltage window.
The mean local moment magnitude drops sharply.
Subgap peaks develop in the density of states,
which gain weight with increasing $V$.
\item 
Low moment AF-M ($V \gtrsim V_c$):
This is a state with a large current, with small 
magnetic moments surviving 
close to the center of the system. It is an 
inhomogeneous antiferromagnetic metal.
\item 
Current saturation in PM-M, ($ V \gg  V_c$):
The spectral features and the current no longer change with
voltage and the moments become vanishingly small.
\end{enumerate}
Apart from the current, and the spatial behaviour of charge
and spin variables, we provide detailed results on the 
voltage dependence of the local DOS. 
We implement a perturbative scheme to evaluate the spectral
features in the small bias regime.

\section{Model and Method}

\subsection{Model}

We consider a 2D Mott insulator connected to non-interacting leads 
via tight-binding coupling. The leads serve two purpose: 
(i)~the bias is introduced through a chemical potential 
difference between the leads, and (ii)~the leads 
provide dissipative channels for the system to relax. 
The isolated Mott insulator when subjected to a field does not 
thermalise for a significantly long time, and instead gets trapped 
in a prethermalised metastable state\cite{freericks2015}. 
The presence of metallic leads at the two edges allows 
relaxation  into a nonequilibrium steady state (NESS)
\cite{ribeiro2015}.

We model the Mott insulator (``system'') by the repulsive Hubbard model 
defined on a square lattice, while the leads are modeled as
tight-binding electron reservoirs. 
Each site at the left and the right edges of the Mott insulator,
Fig.1, 
is coupled to the nearest bath site via tight-binding coupling.
\begin{eqnarray}\label{eq:H}
\mathcal{H}~~~ &=& \, \mathcal{H}_{sys} + \mathcal{H}_{bath} 
+ \mathcal{H}_{coup} \cr
\cr
\mathcal{H}_{sys} &=& \sum\limits_{\substack{<ij>,\sigma}}
-t_{ij} d^{\dagger}_{i\sigma}d_{j\sigma} 
+ U\sum\limits_{i} n_{i\uparrow}n_{i\downarrow} \cr
\mathcal{H}_{bath} 
%
%
&=& \sum_{<ij>,\sigma}t_{\beta}
\left(c^{\dagger\beta}_{i\sigma}c^{\beta}_{j\sigma} + h.c.\right)
-\sum_{\beta,i,\sigma}\mu_{\beta}n^{\beta}_{i\sigma} \cr
\mathcal{H}_{coup} &=& \sum_{<ij>,\sigma}v_{ij}
\left(c^{\dagger L}_{i\sigma}d_{j\sigma} + c^{\dagger R}_{i\sigma}
d_{j\sigma} + h.c.\right) 
\end{eqnarray}
where $t_{s}$, $\mu$ and $U$ are the nearest-neighbour hopping
amplitude, chemical potential and onsite Coulomb repulsion strength,
respectively, in the system. 
$t_{\beta}$ and $\mu_{\beta}$ are the hopping strengths and the
chemical potentials in the metallic baths, where $\beta=(L,R)$,
with $L$ being the left lead and $R$ the right lead.
$\mu_{L,R} = \mu\pm\left(V/2\right)$, V being the applied bias.
$v_{ij}$ denote the matrix elements of the system-bath coupling.

\begin{figure}[t]
\centerline{
\includegraphics[width=7.5cm,height=4cm]{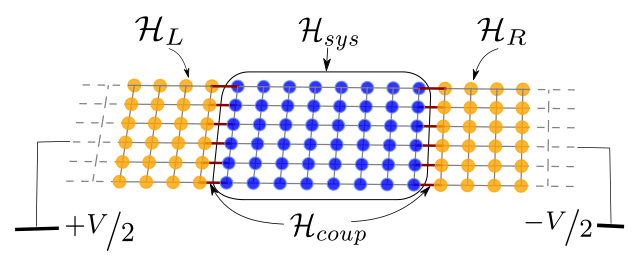}
}
\caption{Schematic diagram showing the setup. The sites having onsite
Hubbard repulsion are marked in blue, while the noninteracting bath
sites are marked in yellow. The coupling between the system and the
bath is denoted by red bonds. The bias is applied symmetrically at the
two edges by tuning the chemical potential of the baths.}
\end{figure}
\begin{figure}[b]
\centerline{ 
\includegraphics[width=\linewidth]{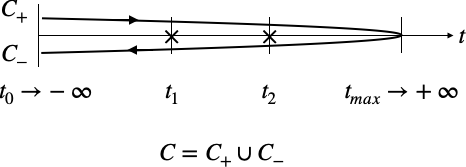}}
\caption{The complex time Keldysh contour for acessing steady states.
In order to access steady states one can take $t_{0}$ (initial time)
and $t_{max}$ (maximum time) to $\pm\infty$. In order to calculate
observables, e.g. the two-point function, one makes insertions at
intermediate times $t_{1}$ and $t_{2}$.}
\end{figure}

\subsection{Method}

\subsubsection{Keldysh formulation}

Assuming the leads were connected far in the past, one can write the 
steady-state action for the system by discretizing the complex time
Keldysh contour as shown in Fig.2. The generating functional
is given by:
\begin{equation}
 Z = \int\mathcal{D}\{\bar{c},c;\bar{d},d\} \, 
e^{i S\left[\bar{c},c;\bar{d},d\right]}
\end{equation}
where $\left(\bar{c},c\right)$ and $\left(\bar{d},d\right)$ 
are the Grassmann fields for the lead and system fermions 
respectively. $S\left[\bar{c},c;\bar{d},d\right]$ is the complex 
time Keldysh action defined on the contour.
\begin{eqnarray}\label{eq:S}
S~~~ &=& \int^{\infty}_{-\infty}
\mathrm{d}t
\left[\mathcal{L}_{sys}(t) + \mathcal{L}_{bath}(t) + \mathcal{L}_{coup}(t)
\right]
\cr
\cr
\mathcal{L}_{sys}(t) &=& \sum_{<ij>}^{\sigma,s}s  
\bar{d}^{s}_{i\sigma}(t)\left(i \partial_{t}+t_{ij}\right)
d^{s}_{j\sigma}(t) -U\sum\limits_{i,s}s \, n^{s}_{i\uparrow}
n^{s}_{i\downarrow}
\cr 
\mathcal{L}_{bath}(t) &=& \sum_{<ij>}^{\sigma,\alpha,s}s  
\bar{c}^{s}_{i\sigma\alpha}(t)\left(i \partial_{t}+t^{\alpha}_{ij}\right)
c^{s}_{j\sigma\alpha}(t)
\cr
\mathcal{L}_{coup}(t) &=& \sum_{<ij>,\sigma,\alpha}v_{ij}
\bar{c}^{s}_{i\sigma\alpha}(t)d_{j\sigma}(t) + h.c.
\end{eqnarray}
where $i$,$j$ are the lattice indices, $\sigma$ is the spin index, 
$\alpha$ labels the leads and $s$ labels the contour. $s$ = $\pm$1 
for the upper and the lower contour fields respectively.

The action is quartic in the system fermion fields. To make progress
we `decouple' the interaction term by introducing a pair of auxiliary
fields at each space-time point.
We proceed as follows:
\begin{subequations}\label{eq:hs}
\begin{equation}
U n^{s}_{i\uparrow}n^{s}_{\downarrow} = \frac{U}{4}
\left(n^{s}_{i}\right)^{2} - U\left(\bm{S}^{s}_{i}\cdot
\bm{\hat{\Omega}}^{s}_{i}\right)^{2}\tag{\ref{eq:hs}}
\end{equation}
\begin{equation}
e^{-i\frac{sU}{4}\left(n^{s}_{i}(t)\right)^{2}} \propto \int\mathrm{d}
\phi^{s}_{i}(t) \, e^{i \left(\frac{sU}{4}(\phi^{s}_{i}(t))^2
- \frac{sU}{2}\phi^{s}_{i}(t)n^{s}_{i}(t)\right)}\label{eq:hs_phi}
\end{equation}
\begin{align}
&e^{isU\left(\bm{S}^{s}_{i}(t) \cdot \hat{\bm{\Omega}}^{s}_{i}(t)\right)^{2}}
\propto\nonumber\\ &\int\mathrm{d}^{3} \bm{M}^{s}_{i}(t)\,
e^{i\left(-\frac{sU}{4}(|\bm{M}^{s}_{i}(t)|)^2
+ sU\bm{M}^{s}_{i}(t)\cdot\bm{S}^{s}_{i}(t)\right)}\label{eq:hs_m}
\end{align}
\end{subequations}
In the equations above, 
$n^{s}_{i} = \bar{c}^{s}_{i\uparrow}c^{s}_{i\uparrow} +
\bar{c}^{s}_{i\downarrow}c^{s}_{i\downarrow}$ 
is the local density, $\bm{S}^{s}_{i} = \frac{1}{2}\sum
\limits_{\alpha\beta}\bar{c}^{s}_{i\alpha}\bm{\sigma}_{\alpha\beta}
c^{s}_{i\beta}$
is the electron spin operator and $\bm{\hat{\Omega}}^{s}_{i}$ is an 
arbitrary SO(3) vector.  Hubbard-Stratonovich 
(HS) transformation of the first term in the interaction introduces
a scalar 
``charge field'' $\phi_{i}(t)$.
The HS transformation on the second term brings in an
auxiliary O(3) ``spin field'' $\bm{M}_{i}(t)$.

As a result of these transformations the problem gets mapped to a 
quadratic Lagrangian with additional fluctuating auxiliary fields 
$\phi^{\pm}_{i}\left(t\right)$ and $\bm{M}^{\pm}_{i}
\left(t\right)$.
We now implement the Keldysh rotation for the fermionic and 
the auxiliary fields, and then integrate out the fermions (system 
as well as leads). The details are provided in the
Appendix\ref{appendixA}. This leaves us with an action which is
dependent on the `classical' and `quantum' auxiliary fields,
which are defined as:
\begin{eqnarray}\label{eq:kelrot}
\phi^{c}_{i}\left(t\right) &\equiv& 
{1 \over 2} 
(\phi^{+}_{i}
\left(t\right) + \phi^{-}_{i}\left(t\right)) \cr
\phi^{q}_{i}\left(t\right) &\equiv& \phi^{+}_{i}\left(t\right) 
- \phi^{-}_{i}\left(t\right)
\end{eqnarray}
for the $\phi$ fields, and similarly for the $\bm{M}$ fields.

\subsubsection{Static Path Approximation (SPA)} 

Till this point the formulation has been exact. We have traded off the 
quartic interaction term for a quadratic theory with four fluctuating 
auxiliary fields. In order to make further progress, we introduce the 
static path approximation, in which we retain only the zero frequency 
mode of the auxiliary fields and ignore their finite 
frequency modes. This would be a drastic approximation if we were to 
consider the transient response of the system. But once 
the system relaxes to a NESS, one expects the average long time
behaviour to be reasonably captured by the zero frequency 
mode of the auxiliary fields. The 
effect of the finite-frequency modes can be built back 
perturbatively on top of the zero-mode theory,
we defer that discussion to another paper. 
Let us introduce the notation:
$$
\phi^{c,q}_{i}(\omega = 0) \equiv \phi^{c,q}_{i},~~~
\bm{M}^{c,q}_{i}(\omega = 0) \equiv \bm{M}^{c,q}_{i}
$$

Under these assumptions we arrive at an effective steady state 
description given by the `static path' action:
\begin{subequations}
\begin{align}\label{S_SPA}
S^{SPA}\left[\phi,\bm{M}\right] &= -i Tr\ln\left[i \check{G}^{-1}(\omega)
\right] + S^{\prime}\left[\phi,\bm{M}\right]
\end{align}
where,
\begin{align}\label{G_SPA}
\check{G}^{-1}\left(\omega\right) \equiv \begin{bmatrix}
\left(\hat{G}^{-1}(\omega)\right)^{R} & \left(\hat{G}^{-1}(\omega)\right)^{K}\\
\hat{0} & \left(\hat{G}^{-1}(\omega)\right)^{A}
\end{bmatrix} + \hat{\Sigma}\otimes\sigma_{x}
\end{align}
with
\begin{align}
\left[\left(\hat{G}^{-1}(\omega)\right)^{R}\right]_{ij}^{\alpha
\alpha^{\prime}} &= \left((\omega+i\eta)\delta_{ij}+\hat{\Gamma}^{R}_{ij\alpha}(\omega)
\right)\delta_{\alpha\alpha^{\prime}}-\hat{H}_{ij}^{\alpha\alpha^{\prime}}\\
\left[\left(\hat{G}^{-1}(\omega)\right)^{A}\right]_{ij}^{\alpha
\alpha^{\prime}} &= \left((\omega-i\eta)\delta_{ij}+\hat{\Gamma}^{A}_{ij\alpha}(\omega)
\right)\delta_{\alpha\alpha^{\prime}}-\hat{H}_{ij}^{\alpha\alpha^{\prime}}\\
\left[\left(G^{-1}(\omega)\right)^{K}\right]_{\substack{ij;\\\alpha
\alpha^{\prime}}} &=\hat{\Gamma}^{K}_{ij\alpha}(\omega)
\delta_{\alpha\alpha^{\prime}} + \nonumber\\
F(\omega-\mu)&\left[\left(\hat{G}^{-1}(\omega)\right)^{R}-
\left(\hat{G}^{-1}(\omega)\right)^{A}\right]_{ij}^{\alpha\alpha^{\prime}}
\\
\Gamma^{R,A,K}_{ij\alpha}\left(\omega\right)=&\sum_{\beta\in\{L,R\}}
\left(\sum_{mn}v_{mi}v_{nj} g^{R,A,K}_{\beta,mn\alpha}(\omega)\right)\\
\hat{H}_{ij}^{\alpha\alpha^{\prime}} = -t_{<ij>}&\delta_{\alpha\alpha^{\prime}}
+ \frac{U}{2}(\phi^{c}_{i}\delta_{\alpha\alpha^{\prime}} - \bm{M}^{c}_{i}\cdot
\bm{\sigma}_{\alpha\alpha_{\prime}})\delta_{ij}\label{H}\\
\hat{\Sigma}_{ij}^{\alpha\alpha^{\prime}} = -\frac{U}{4}&(\phi^{q}_{i}
\delta_{\alpha\alpha^{\prime}} - \bm{M}^{q}_{i}\cdot
\bm{\sigma}_{\alpha\alpha_{\prime}})\delta_{ij}
\end{align}
and
\begin{align}
S^{\prime}\left[\phi,\bm{M}\right] = \frac{U}{4\pi}\sum_{i}
\left(\phi^{c}_{i}\phi^{q}_{i}-\bm{M}^{c}_{i}\cdot\bm{M}^{q}_{i}\right)
\end{align}
\end{subequations}
$g^{R,A,K}_{\beta}(\omega)$ denote the retarded, advanced and Keldysh components
of Green's function of the reservoirs, while $\hat{G}^{R,A,K}$ denote those
of the system. $\hat{\Gamma}^{R,A,K}$ is the dissipation introduced
in the system due to the leads. $\hat{H}$ is an effective Hamiltonian which
one can obtain if one retains only the zero frequency mode of the auxiliary fields.
$F(\omega-\mu)$ is the distribution function of the isolated system.
$\eta$ is a small positive number, which regulates the
Keldysh action.

The mean-field consistency conditions can be obtained by extremising 
the SPA action with respect to the quantum auxiliary fields. 
We get the following family of saddle point equations:
$$
\frac{\delta S^{SPA}}{\delta\phi^{q}_{i}}\at[\bigg]
{\phi^{q},\bm{M}^{q}=0} = 0,~~~
\frac{\delta S^{SPA}}{\delta \bm{M}^{q}_{i}}\at[\bigg]
{\phi^{q},\bm{M}^{q}=0} = 0
$$
These can be simplified to obtain the consistency conditions:

\begin{subequations}\label{MF:main}
\begin{align}
\int\mathrm{d}\,\omega\, \Im\left[Tr\left(\hat{G}^{K}_{ii}(\omega)\right)
\right]~~ &=
~\phi^{c}_{i}\label{MF:phi}\\
\int\mathrm{d}\,\omega\, \Im\left[Tr\left(\hat{G}^{K}_{ii}(\omega)
\bm{\sigma}\right)\right] 
&= 
~\bm{M}^{c}_{i}\label{MF:M}
\end{align}
\end{subequations}
where $\hat{G}^{K}(\omega) = -\hat{G}^{R}(\omega)
\left(\hat{G}^{-1}(\omega)\right)^{K}\hat{G}^{A}(\omega)$
and the trace is over the $2\times 2$ spin subspace.

\begin{figure}[t]
  \centering
  \includegraphics[width=\linewidth]{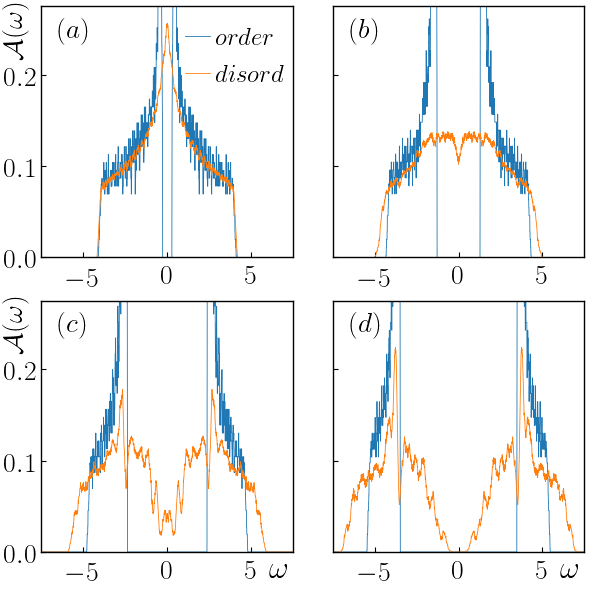}
  \caption{(a)-(d) Comparing the density of states in the antiferromagnetic
    state and the `paramagnetic' phase with random moment orientations for
    $U/t = 2,4, 6, 8$ respectively. The ordered state remains gapped
    at all values of $U$. At $U/t = 2$, loss of order creates a gapless DOS.
    At $U/t = 4$, the `paramagnetic' state is gapless but band
    singularities are absent. At $U/t = 6$ there is a pseudogap,
    while $U/t = 8$ shows a clean gap persisting in the `paramagnetic'
    state.}
  \label{fig:dos_cmp}
\end{figure}

\begin{figure}[b]
  \centering
  \includegraphics[width=\linewidth]{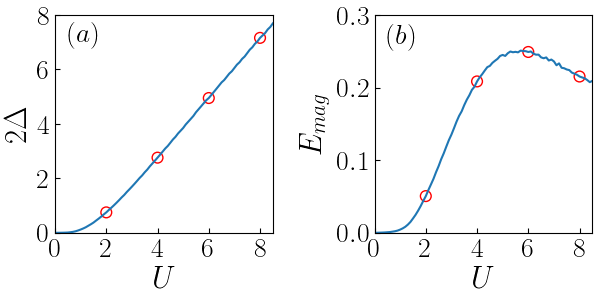}
  \caption{(a) Charge gap $2\Delta$ as a function of $U$ for the
    Hartree-Fock ground state at equilibrium. (b) Energy difference
    $E_{mag}$ between the antiferromagnetic ground state and the
    `paramagnetic' state with random moment orientations, as a
    function of $U$. $E_{mag}\sim t^{2}/U$ at large U.}
  \label{fig:emag}
\end{figure}

\begin{figure}[b]
\centerline{
\includegraphics[width=4.5cm,height=4.5cm]{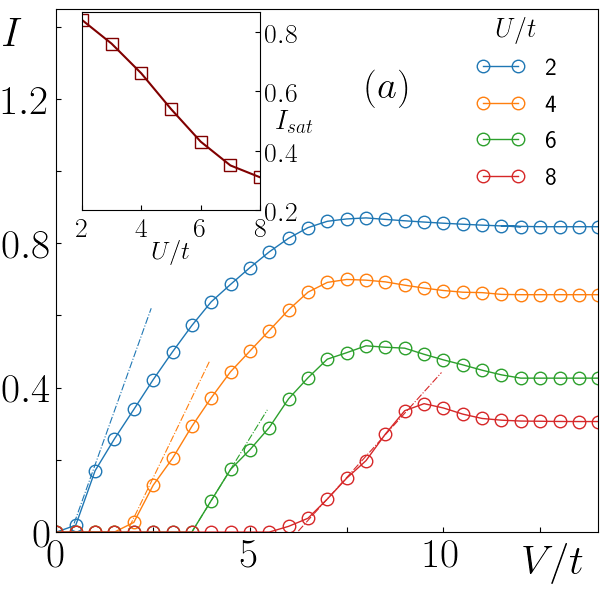}
\includegraphics[width=4.5cm,height=4.5cm]{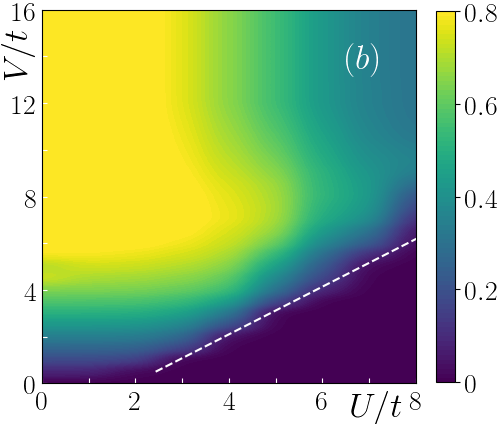} }
\caption{(a)~The $I-V$ characteristics for different values of
interaction strength $U$, showing the transition from an insulating
Mott state to a metallic state. The current rises sharply around
a `critical bias' $V_{c}$, and finally saturates to a scale $I_{sat}(U)$
when $V \gg V_c$.  The dependence of $I_{sat}$ on the $U$ is shown in
the inset.  (b)~A map of the current for varying $U$ and $V$. The broken
line demarcates the insulator-metal `phase boundary'.
} \end{figure}

\begin{figure}[t]
\centerline{
\includegraphics[width=7.6cm,height=4.5cm]{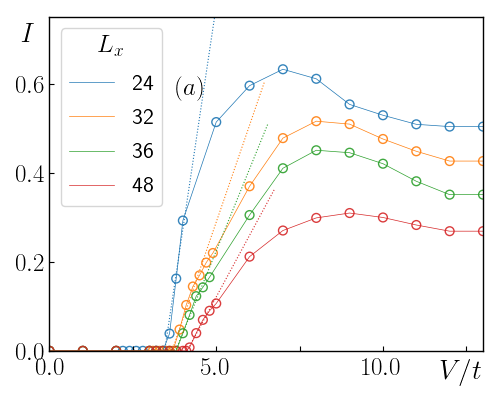}
}
\centerline{
~~
\includegraphics[width=3.8cm,height=4cm]{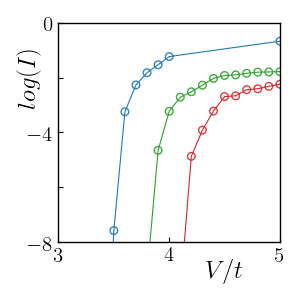}
\includegraphics[width=3.8cm,height=4cm]{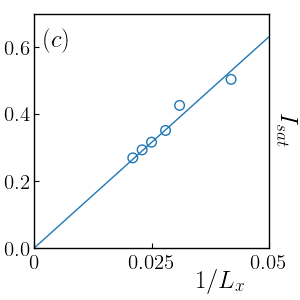}
}
\caption{(a)~Size dependence of $I-V$ characteristics at $U = 6$. Both
the critical bias $V_{c}$, as well as the saturation current $I_{sat}$
depend on the longitudinal size of the system.  (b)~The region around
$V_{c}$ has been plotted in log scale to highlight its size dependence.
(c)~Shows the dependence of $I_{sat}$ on the longitudinal size. The data
can be fitted reasonably to $I_{sat} \propto 1/L_x$.
For a fixed $V$ the current, and the `metallisation' effect in
general, would vanish as $L_x \rightarrow \infty$.}
\end{figure}

\begin{figure*}[t]
\centerline{
  \includegraphics[width=18cm,height=8.0cm]{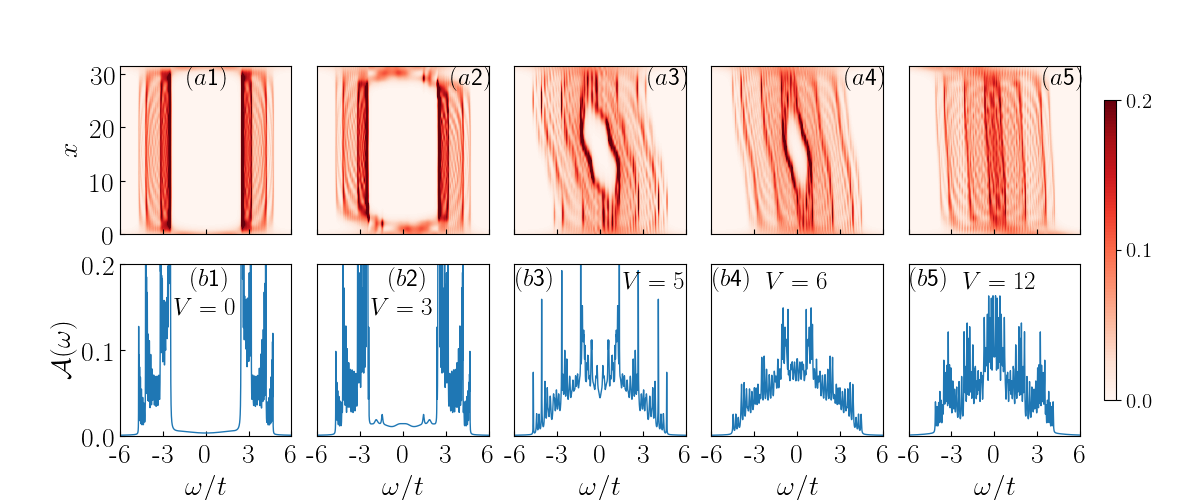} }
\caption{
Density of states: the upper row
shows the variation of local DOS $A_{ii}(\omega,x)$ along
the longitudinal direction, $(x)$, with changing bias at $U = 6t$.
The panels a1-a5 are for $V = \{0, 3, 5, 6, 12\}$ respectively.
The regimes are
(a1)~the reference AF-I state at $V=0$, (a2)~AF-I at low $V$, 
(a3)~AF-I to AF-M breakdown, (a4)~AF-M at large current, 
(a5)~PM-M state with current saturation. 
The lower row shows the behavior of the system averaged 
density of states with changing bias at $U = 6t$. Panels b1-b5 
are for the sames values of $V$ as the respective panel above.}
\end{figure*}

\subsubsection{Implementation of mean field consistency}

Our method involves treating the Hubbard interaction within an
unrestricted Hartree-Fock (HF). We have benchmarked the scheme
against the analytically tractable zero temperature mean-field
limit of the equilibrium $(V=0)$ Hubbard model in 2D with
periodic boundary conditions. Starting with an initial guess
for $\{\phi^{c} \equiv \phi\}$ and $\{M^{c}_{z}\equiv M\}$,
we calculate the auxiliary fields via frequency integrals
of the locally projected Green's functions. The mean field
consistency equations are solved iteratively until the solutions converge
for every site. The convergence criteria used is $max_{i}\left[
\left(|\Vert{\bf M}\Vert^{n+1}_{i}-\Vert{\bf M}\Vert^{n}_{i}|\right)
/\Vert{\bf M}\Vert^{n}_{i}\right] <= 0.01$
or $max_{i}\left[|\Vert{\bf M}\Vert^{n+1}_{i}-\Vert{\bf M}\Vert^{n}_{i}|\right] < 10^{-6}$, where $\Vert{\bf M}\Vert^{n}_{i}$ is the magnitude of the
local moment at site $i$ in the $n$-th iteration. A similar criteria 
has used for the $\phi$ field as well.

In principle, one must solve for all the three components of $\bm{M}^{c}
$ at each site. For the square lattice, one can simplify the problem by 
retaining only the $M^{c}_{z}$ component of the spin auxiliary field. 
This can be justified by performing a strong coupling expansion of the 
Keldysh action to find that the collinear state remains a consistent 
solution even at finite bias. We have elaborated on
this point in Appendix\ref{appendixB}.

\subsubsection{Nature of the $V=0$ insulating state}
\label{sec:cmp}

Before launching into the voltage response we quickly
discuss the nature of the $V=0$ state.
The ground state of the half filled Hubbard model
on a square lattice with nearest neighbour hopping is
an antiferromagnetic insulator (AF-I) for
all values of $U/t$. However, there is a qualitative
difference between the $U/t \lesssim 1$ insulator and the
$U/t \gg 1$ insulator. For  $U/t \lesssim 1$  one
obtains a `Slater insulator' where the charge gap is
associated with the magnetic order. When $U/t \gg 1$, however,
the charge gap (or a pseudogap) survives even when magnetic
order is lost. This is the `Mott insulator'. 

To locate the Slater to Mott crossover we used two indicators.
In the first, Fig.\ref{fig:dos_cmp}, we compare the density of
states (DOS) in the AF-I ground state with that in a 
`paramagnetic phase' - where the moment magnitude
is same as the AF-I state but the orientations are 
randomised.  The results show that at 
$U=6t$ the loss of magnetic order still leaves
a prominent pseudogap. At this $U$ the $V=0$ state
is a Mott insulator.

The second indicator is related to charge and spin excitation
energies.  The charge gap $2\Delta$ between the upper and
lower Hubbard bands at $T = 0$ crudely defines the 
temperature scale
$T_{gap}\sim 2\Delta \sim U$ (at large $U$) at which the DOS
would become gapless. 
Another scale  $E_{mag}$ arises from the energy difference
between the perfect AF-I state and the random orientation 
state. This defines the temperature scale at which 
magnetic order would be lost.
Fig.\ref{fig:emag} shows that for $U/t =6$ we are in a 
regime where $2 \Delta \gg E_{mag}$. In fact $2 \Delta \sim
U$, while $E_{mag} \sim t^2/U$. Again the signature of a
Mott insulator.

A full thermal theory has indeed been developed for the
equilibrium problem  \cite{anamitra,rajarshi,nyaya} and
recovers $T_c$ scales consistent with Fig.\ref{fig:emag}.
See Fig.1 in \cite{anamitra}. We are developing a non
equilibrium generalisation of this approach\cite{finiteT}.

\section{Results}

We discuss the results for our implementation of the scheme in a 
2D system, which is finite in the longitudinal direction, while being
periodic in the transverse direction. 
For the tight-binding baths, we approximate the
density of states by an appropriate Lorentzian function.
\begin{equation}
\rho_{L,R}(\omega) = \left(\frac{D}{3 \tan^{-1}(\frac{3}{2})}\right)
\frac{1}{\left(\omega^{2}+\left(\frac{2D}{3}\right)^{2}\right)}
\end{equation}
  Where $D$ is the bandwidth of the bath.
After integrating out the bath we retain only the diagonal terms arising from
the bath Green's functions, which are proportional to the bath DOS,
an approximation that is justified in the wide-bandwidth limit \cite{rubtsov2016}.

Unless explicitly mentioned, we have shown the results for a system with 32
sites in the longitudinal (x) direction and 8 sites in the transverse (y) 
direction. We have also studied the size dependence of our results
by varying the longitudinal size from 12 to 48 sites, and the transverse
size from 4 to 8 sites. The size dependence has been discussed in the 
relevant sections. Unless otherwise mentioned, all energies are measured 
in units of hopping in the system $t_{s}$, and all currents are 
measured in units of $2et_{s}/{\hbar}$.

\subsection{I-V character and `phase diagram'}

The bond current between the nearest neighbour sites
in the $x$ direction is given by the expression:
\begin{eqnarray}
I_{j,\,j+1}\left(V\right) &=& \sum_{\sigma}\int_{-\infty}^{\infty}
\frac{\mathrm{d}\,\omega}{2\pi}\left(G^{<}_{j+1,\,j;\sigma}(\omega)
-G^{<}_{j,\,j+1;\sigma}(\omega)\right)
\cr
G^{<}(\omega)  &=&  \frac{1}{2}\left(G^{K}(\omega) +
G^{A}(\omega) - G^{R}(\omega)\right) 
\nonumber
\end{eqnarray}
For the system in consideration, which has periodic boundary conditions
along the transverse direction, the current is found to be the same
along all longitudinal chains. Hence, the total current scales linearly
with the transverse size of the system.
At steady state one further expects the current to be the same on all 
bonds along the longitudinal axis. This has been numerically checked to be
true strictly in the limit $\eta\rightarrow 0$.

For a sufficiently large system size and for $U \gtrsim 2$
(that we have explored given our size constraints) 
we find that I-V characteristic
has three different regions:
(i)~ pre-breakdown - exponentially suppressed current,
(ii)~breakdown - the current increases rapidly to attain it's
maximum value, and (iii)~saturation - the current drops from the
maximum to saturate at a finite value.
Regime (ii) involves a transition from  an AF-I to an AF-M,
both spatially inhomogeneous, as we have stated earlier.

In Fig. 5(a) we plot the current $\left(I\right)$ 
as a function of bias $\left(V\right)$ for various interaction 
strengths $U$. In regime (i), when $V \ll V_c$,
the current $I\sim e^{-V_{c}/V}$, Fig. 6(b), 
after which the system enters regime
(ii), where the current increases rapidly, and then
saturates in regime (iii) where $V \gg V_c$. 
For $U > 2$, and moderate sizes we encounter a region of negative 
differential resistance (NDR) in the I-V between the breakdown
and saturation regimes. On studying the size dependence [Fig. 6(a)]
at a fixed value of $U$, one finds that this effect diminishes with
increasing length of the transport direction $L_x$, and ultimately
goes away at large enough sizes. One has encountered such regions of
NDR in experiments\cite{zimmers}.

The saturation current $I_{sat}$ depends on the length of the
system in the direction of transport $L_x$ [Fig. 6(a)]. Being formulated
in real space, our method allows us to study the size dependence of
the saturation current. As shown in Fig. 6(c), $I_{sat} \sim 1/L_x$,
vanishing as $L_{x}\rightarrow\infty$.

One can construct a `phase diagram' from the current
map, $I(U,V)$, in Fig. 5(b). 
It shows the required threshold bias as a function of $U$
for a fixed size of the system, $V_{c}\left(U,L_{x}\right)$. Within our
calculation we find $V_{c}\left(U,L_{x}\right)$ to be dependent on
the longitudinal size of the system. This too has been observed in
experiments\cite{zimmers}. The phase diagram separates the
insulating regime (i) from the metallic regimes (ii) and (iii),
in the $U-V$ plane. Note that for the square lattice, at half filling,
the ground state becomes antiferromagnetic for an arbitrarily small $U$,
but for $U/t \gtrsim 4$ the insulating character can survive even in the
absence of magnetic order. 

Within our scheme the Mott insulator supports
exponentially weak current at $V < V_{c}\left(U\right)$
[Fig. 6(b)], but beyond $V_{c}\left(U\right)$ the
current rises
sharply and then reduces towards a saturation value.
This is in contrast to the DMFT\cite{okamoto2008,eckstein2010,aron-prl2012} and
Gutzwiller mean field\cite{mazza2015} studies in which the current
appears to have a functional form $I \sim V e^{-\alpha/V}$.

The $I-V$ curve in our
theory has a point of inflection (second derivative vanishes) at the $V_{c}$,
while the DMFT form does not have any such point. Such a point arises within
our theory because the current saturates at large $V$, while the DMFT current
keeps growing linearly at large $V$. Within our theory, the current saturation
occurs due to the finite bandwidth of system, irrespective of the bandwidth of
the bath.  The behaviour we observe has similarity to some
experiments\cite{kumai1999, nakamura2013}.

\subsection{Density of states}

The density of states (DOS), $\mathcal{A}\left(\omega\right)$, 
is obtained by averaging the local DOS, $\mathcal{A}_{ii}(\omega)$,
over the $x$ direction.
\begin{subequations}
\begin{align}
  \mathcal{A}_{ii}\left(\omega\right) = \sum\limits_{\sigma}\mathcal{A}_{ii,\sigma}
  &= -\frac{1}{\pi}\sum\limits_{\sigma}Im\left(G^{R}_{ii,\sigma}
    \left(\omega\right)\right)\\
\mathcal{A}\left(\omega\right) &= \frac{1}{L}\sum\limits_{i=1}^{L}
\mathcal{A}_{ii}\left(\omega\right)
\end{align}
\end{subequations}
\begin{figure}[b]
\centerline{
  \includegraphics[width=4.2cm,height=3.5cm]{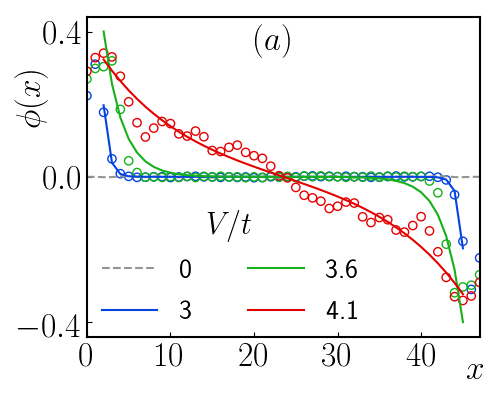}
  \includegraphics[width=4.2cm,height=3.6cm]{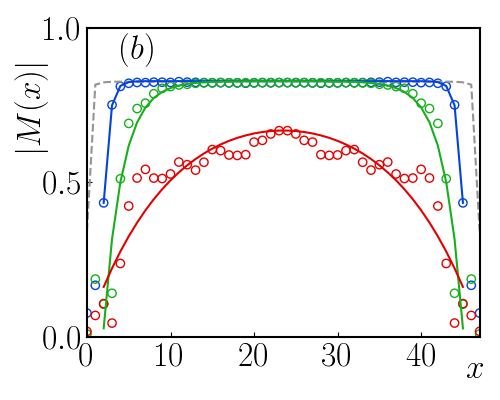} }
\centerline{
  \includegraphics[width=4.2cm,height=3.5cm]{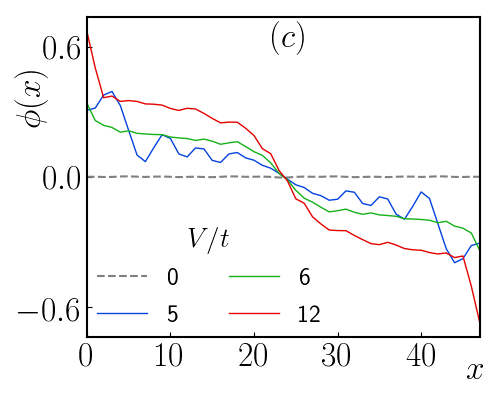}
  \includegraphics[width=4.2cm,height=3.6cm]{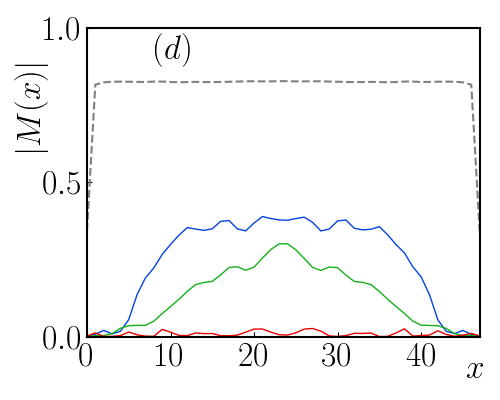} }
\caption{ Top: The auxiliary fields, $\phi(x,V)$ and $|M(x,V)|$
as a function of bias for $U = 6t$. Panels $(a)$ and $(b)$ show
the spatial profile for a few values of $V$ corresponding to the
pre-breakdown regimes in the $I-V$ curve. The open circles denote
the data points, while solid lines denote fitted curves for
$\phi$ in $(a)$ and $|M|$ in $(b)$
using the trial functions. The broken lines
in grey which denote the zero bias profiles for the auxiliary
fields serve as the reference.  Bottom:
The auxiliary field profiles in the post-beakdown $V = 5,6$
and saturation $V = 12$ regimes.  }
\end{figure}

\begin{figure}[t]
\centerline{
\includegraphics[width=5.5cm,height=5.5cm]{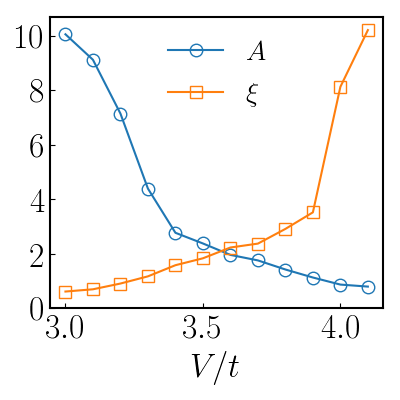}}
\caption{ The fitting parameters - penetration length $\xi$ and scale
factor $A$ as a function of bias $V$, in the neighbourhood of
the crossover. The results shown are for a $48\times 8$ system
at $U = 6$.  }
\end{figure}

\begin{figure*}[t]
\centerline{
\includegraphics[width=15.2cm,height=7.1cm]{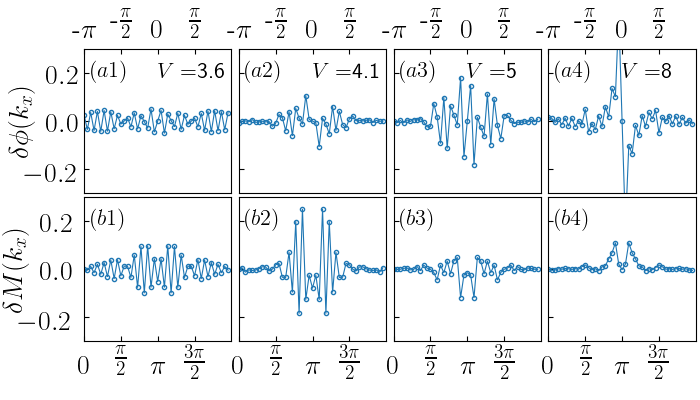}
}
\caption{Fourier content of the auxiliary field modulations at 
$U = 6t$ plotted for different $V$. The upper panels show the Fourier
transform of $\delta\phi(x) = \phi(x) - \phi_{tr}(x)$, while the
lower panels show the Fourier transform of $\delta M(x) = M(x) - M_{tr}(x)$.
Voltages are $V = \{3.6, 4.1, 5, 8\}$ from left to right columnwise.
As $V$ increases across the transition, the interval containing the
dominant modes shrinks to a narrow interval around $k_{x} = 0$ for
$\delta\phi(k_x)$ and $k_{x} = \pi$ for $\delta M(k_x)$ respectively.
$L_{x} = 48$ for the plot shown.  
}
\end{figure*}

At $V=0$ the LDOS is gapped at all sites,
except for a couple of sites at the edges
- which get renormalised due to coupling 
with the metallic baths. The sites away from the edges remain
essentially unaffected in the absence of bias, as is evident 
in Fig. 7(a1). The system averaged  DOS,
Fig. 7(b1), remains largely indistinguishable from that of an 
isolated system.
At low bias, $V \ll V_c$, 
the effect of bias decays exponentially 
inside the system, as we will see in the next section, 
with a `penetration length' controlled 
by the bias. As the penetration length increases,
sites progressively away from the edges start to `sense' the bias
and the gap in the LDOS collapses for these sites, Fig. 7(a2).
One can now find subgap state in the DOS, even though the edges
in the global DOS, Fig. 7(b2),
remain sharply defined. In this regime the behaviour of
the DOS can be understood, approximately, in terms of a
weak site-dependent
perturbation on top of a translation invariant parent state. We
discuss this further in Sec.\ref{pert}.
 
After the breakdown, $V \gtrsim V_c$, 
all the sites feel the effect of the bias and a significant fraction
of sites towards the edges become gapless, while the gap for sites
at the center gets suppressed significantly [Fig. 7(a3)]. 
The overall DOS [Fig.7 (b3)]
becomes gapless indicating that the fraction of ungapped sites
is comparable to the gapped ones. In this regime, the gap in the LDOS
shrinks with increasing $V$ [Fig. 7(a4)], and ultimately vanishes as we
hit the saturation regime [Fig. 7(a5)]. The DOS develops
increasing weight around $\omega =0$.

\subsection{Charge and spin profiles}

Within our framework the effect of the bias gets 
encoded in the auxiliary fields through the
 self-consistency. The electrons respond to the 
auxiliary field patterns and sense the bias 
through them. For the unbiased open system the 
charge field $\phi$ vanishes throughout the system, 
whereas the spin field magnitude is uniform  
($\leq 1$) on all sites, except for the edge sites.
We have checked that the spin field alternates from site to 
site even at finite $V$.

If the bias is applied symmetrically, i.e. $\mu_{L}+\mu_{R} = 2\mu$
condition is satisfied, then the auxiliary fields must have a
specific symmetry: 
the charge field $\phi_i$ is
antisymmetric with respect to the center of the system,
while $\vert {\vec M}_i \vert$ is symmetric. 
Fig.8(a) and 8(b) (open circles) show the bias dependence of the $\phi$  
and the $|M|$ fields in the pre-breakdown regime at 
$U = 6t$.
Till close to $V_{c}$ the effect of the bias
remains confined to the edges.
In this regime, we obtain a ``mean profile'' for the auxiliary fields by
fitting them with the trial functions:

\begin{subequations}
\begin{align}
    \phi_{tr}\left(x\right) &= A\left(e^{-x/\xi} -
    e^{-(L_{x}-x)/\xi}\right)\\
    M_{tr}\left(x\right) &= \left(M_{0}- A\left(e^{-x/\xi} +
    e^{-(L_{x}-x)/\xi}\right)\right)\times cos\left(\pi x\right)
\end{align}
\end{subequations}
where $L_{x}$ is the length in the longitudinal direction. 
$M_{0}$ is the mean-field moment size for the translation
invariant system.
The bias dependence of the deviation fields enters 
through the parameters $A$ - an overall scale factor, 
and $\xi$ - the penetration length. 
The fitted curves for $\phi$ and $|M|$ have been plotted using
solid lines in Fig.8(a) and 8(b).

Fig.7 shows the bias dependence of fitting
parameters.For $V \ll V_{c}$ the effect of bias remains
localised to the edges and $\phi$ and $|M|$ 
decay exponentially within a few sites from 
the edges. As $V$ approaches $V_{c}$ the effect of 
bias penetrates deeper, indicated by the rapidly
rising penetration length for both the fields. Once the
penetration length attains a significant value, the entire 
system senses the effect of bias and the current rises
rapidly. Across the transition $\xi$ grows continuously by
almost an order of magnitude. 
The scale factor $A$ 
falls rapidly before the transition and then 
saturates to unity beyond the transition, for all system 
sizes. Hence, in terms of the `deviation fields'
the Mott breakdown can be visualised 
as a transition from a strong deviation close
to the edges to a system wide effect near the transition.

Beyond $V_c$ the fields show a spatial
modulation on top of a mean profile, as is evident in Fig.8(c)
and 8(d). The modulations tend to die off as
one approaches the saturation regime. In this regime the mean
$\phi$ acquires a linear profile while the mean $|M|$ becomes
vanishingly small. In order to analyse the mode content on top
of the mean profiles we subtract out the mean curve and then
Fourier analyse the modulating fields $\delta\phi(x)$ and
$\delta M(x)$ in the longitudinal direction, Fig.10.
Before breakdown the auxiliary field profiles are dominated 
by the mean curves and hence, the Fourier profiles of the
deviations are vanishingly small. Even for $V \lesssim V_{c}$,
Figs. 10 (a1) and (b1), one finds that the Fourier weights are
small and evenly spread over the
entire $k$ range. For $V\approx V_{c}$, Figs. 10 (a2),(b2) and
(a3),(b3), the dominant weights lie in an interval of
$\pm\frac{\pi}{2}$ around the $k = 0$ mode $\phi$ and $k = \pi$
mode for $M$. As $V$ increases beyond $V_{c}$, the interval
containing the dominant modes progressively shrinks.
At $V \gg V_{c}$, Figs. 10 (a4) and (b4), the Fourier profile
is dominated by a few modes in the vicinity of $k_{x} = 0$ and
$\pi$ for the charge and spin fields respectively. As noted in
ref.\cite{rubtsov2016}, we find that spatially modulated 
patterns arise in the auxiliary fields as the system metallises,
indicating that the system is susceptible to pattern
formation.
We emphasize that the transition itself is not
driven by these patterns but by the suppression of the
magnetic moments through the bulk of the system.

\section{Discussion}

In what follows we briefly touch upon the nature of the 2D
transition and the large bias state, 
discuss an approximate method for accessing the
$V$ dependence of spectral features, and 
discuss some of the numerical issues related to our
computation.

\subsection{Nature of the transition and large bias state}

Most experiments on the bias driven insulator-metal 
transition show (i)~a first order jump from the insulating to
the conducting state on increasing bias, (ii)~a hysteretic
response to bias cycling, and (iii)~a seemingly
growing current, an `ohmic response', 
as the bias is increased past breakdown.
We do not see these in our results at $U=6t$ so some
clarification is in order

{\it Nature of the transition:} The I-V curves for the bias
driven insulator-metal transition, within this study,
grow smoothly across $V_{c}$, from an exponentially suppressed
current state to high current state. The transition within our
\emph{inhomogeneous} mean field study does not require the
magnetic order parameter to vanish throughout
the system, rather it is driven by a ‘penetration length’
becoming comparable to system size - with moments still
surviving in the center of the system. This is specific to low
spatial dimensions - 1D and 2D, and unlike the scenario reported
in Ref.\cite{han}, where the system undergoes a discontinuous
transition with respect to the applied field. This crucial
difference might be due to the assumption of a uniform order
parameter in that study. However, a 3D generalisation\cite{finiteT}
of this method yields I-V characteristics in which the
current switches discontinuously between the two regimes.
There we also find hysteresis with respect to upward and
downward voltage sweeps. This is the scenario reported in
most of the experiments
\cite{kumai1999,nakamura2013,wang,zimmers,radu,wu} on
voltage driven Mott insulators.

{\it Large bias state:} Within our scheme the current 
saturates at large values of bias since the current
kernel is bounded by the bandwidth of the bath-connected
system. Increasing $V$ beyond this scale does not have any effect on the
current. The range of $V$, post-breakdown, for which the current keeps
growing depends on the ratio of gap ($\Delta$) to the bandwidth ($W$).
For example, in Fig.5(a), for $U/t = 2$ we have $\Delta/W \approx 10$,
and one needs to apply a bias which is ten times larger than
the breakdown voltage 
to find current saturation. This might be the scenario
for most experiments\cite{kumai1999,nakamura2013,wang,zimmers,wu}
where they do not report saturation upto $V \sim 2V_c$.

\begin{figure}[t]
\centering
\includegraphics[height=5cm,width=6cm]{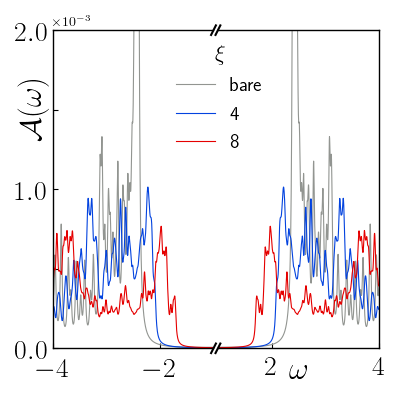}
\caption{The perturbatively corrected DOS, in the pre-breakdown
regime for different values of $\xi$ keeping A fixed at 1.5.
The gap reduces with increasing $\xi$.
}
\label{fig:pertdos}
\end{figure}

\subsection{Spectrum at finite bias, $V \ll V_c$}\label{pert}

We set up a perturbative calculation which captures the gap
renormalisation in the global DOS at finite
bias, $V \ll V_{c}$. In this
regime, the deviation of the auxiliary fields from the
translation invariant background profile is small, and one
can treat the deviation fields $\delta\phi$ and $\delta M$
within perturbation theory. In this, we neglect the contribution
coming from renormalisation of the local DOS at the edges
due to the hybridisation with the leads. This is justified
because the edge contribution to the global DOS is suppressed by
a factor of $1/L_{x} \ll 1$. Under these assumptions, the bare
problem becomes analytically tractable. One can obtain the bare
DOS by diagonalising a $2\times 2$ Hamiltonian in momentum space.
The deviation fields:
\begin{subequations}\label{eq:auxdev}
\begin{eqnarray}
\delta\phi(x) &= Ae^{-L/2\xi}\left(e^{\left(\frac{L}{2}-x\right)/\xi}
                      - e^{-\left(\frac{L}{2}-x\right)/\xi}\right)\\
\delta M(x,y) &= -Ae^{-L/2\xi}\left(e^{\left(\frac{L}{2}-x\right)/\xi}
                + e^{-\left(\frac{L}{2}-x\right)/\xi}\right)\nonumber\\
  &\times\cos(\pi x + \pi y)
\end{eqnarray}
\end{subequations}
are treated within perturbation theory.
We compute the second order self energy corrections to the bare
DOS and show, in Fig.11, that for a fixed $A$, the gap in the
spectrum decreases with increasing $\xi$. The details of the
calculation are provided in Appendix\ref{appendixC}.

\subsection{Numerical checks}

\subsubsection{Size dependence}

Our problem is formulated explicitly
for a finite sized system. For a fixed $V$, if one takes 
$L_{x}\rightarrow\infty$
one does not expect to see a breakdown since the 
field inside the system vanishes.
We have studied the size dependence of the $I(V)$ character in Fig.6.
The gross features in the DOS, namely the gap and the bandwidth, are 
less sensitive to longitudinal size variation.
The fit of the auxiliary field profiles with the
trial functions suggested in Eq.10(a) and 10(b) systematically 
improves with increasing
longitudinal size. Moreover the penetration depth, $\xi$ rises more 
sharply around
$V_{c}$ with increasing $L_{x}$. The transverse size $L_{y}$ 
does not have any effect on the auxiliary field profiles.

\subsubsection{Frequency discretisation} 

In order to stabilise the
Schwinger-Keldysh action, even for an isolated system,
one needs to work with a small, but finite $\eta$. Microscopically, this can be
interpreted as slight broadening of exact eigenlevels due to weak coupling with
the environment. In a numerical implementation, $\eta$ should be chosen to be as
small as allowed by numerical stability, and certainly smaller than the average
level spacing. We find that the choice of $\eta$ can affect the results 
significantly.
The transition
itself gets `smeared' in this scenario. Hence, $\eta$ has been chosen to be much smaller
(an order of magnitude) than the average level spacing and we have ensured that
reducing it further does not affect our results. The choice of $\eta$ also dictates
the frequency discretisation $\Delta\omega$, as one must satisfy the condition
$\Delta\omega \ll \eta$ in order to satisfy the DOS sum rules. This forces us to
work with a very large frequency grid $\sim 5\times 10^{3}$.

\section{Conclusion}

We studied the breakdown of a two dimensional
antiferromagnetic Mott insulator 
in response to a voltage bias, using Keldysh 
mean field theory at zero temperature. We obtained 
the current-voltage characteristics for a finite 
sized system and studied its size dependence. 
The I-V results  which we obtain show a 
threshold behaviour, unlike those obtained 
from dynamical mean field theory studies. 
We studied the variation of the local density of
states in the longitudinal direction in response
to the bias.
The LDOS changes from `gapped' to `ungapped' as 
one moves from the center to the edge at 
$V < V_c$, with the fraction of `ungapped' sites
increasing as one heads towards $V_c$. 
These effects emerge due to the progressive `penetration'
of the applied bias into the bulk over a lengthscale $\xi(V)$, 
which grows to system size near the critical bias.
All response functions can be calculated, approximately,
based on a knowledge of this lengthscale.
The results emphasize the role of spatial symmetry breaking
due to the applied bias, and the need for a real space
treatment of the resulting problem. Our method readily
generalises to disordered and frustrated Mott systems.

~\textit{Acknowledgements}:
We acknowledge the use of HPC Clusters at HRI. 
AD acknowledges fruitful discussions with Abhishek 
Joshi and Sauri Bhattacharyya. AD would also 
like to thank Dibya Kanti Mukherjee and Arpita Saha 
for constructive criticism of the manuscript. 
The research of AD was supported in part by
 the Infosys scholarship for senior students.

\appendix

\section{ Derivation of consistency conditions}\label{appendixA}

Here we present a more detailed derivation of the consistency 
conditions for the static auxiliary fields. Stating from the model 
defined by the Hamiltonian in eq.\ref{eq:H} one can construct the 
generating functional by discretizing the complex time contour as 
given in eq.\ref{eq:S}. Upon introducing the auxiliary fields 
for the upper and the lower contours at each time slice, as given in 
eqns. \ref{eq:hs_phi} and \ref{eq:hs_m}, the partition function can 
be written as:
\begin{subequations}\label{Z_aux:main}
\begin{equation}
Z \propto \int\mathcal{D}\{\bar{c},c;\bar{d},d;\phi,M\}\,
e^{i  \tilde{S}[\bar{c},c;\bar{d},d;\phi,M]}
\end{equation}
with
\begin{align}
\tilde{S} &= \int\limits^{\infty}_{-\infty}\mathrm{d}t\,
\left(\tilde{\mathcal{L}_{S}}(t) + \mathcal{L}_{B}(t) 
+ \mathcal{L}_{C}(t)\right)\label{S_aux}\\
\tilde{\mathcal{L}_{S}}(t) &= \sum\limits_{\substack{i,j\\\sigma,s}}s
\,\bar{d}^{s}_{i\sigma}(t)\left(i \partial_{t}-\phi^{s}_{i}
\delta_{ij}+t_{<ij>} + \sigma_{z}M_{i}\delta_{ij}\right)
d^{s}_{j\sigma}(t)\nonumber\\
&+ \frac{1}{U}\sum\limits_{i,s}\left(
\left(\phi^{s}_{i}(t)\right)^{2}-\left(M^{s}_{i}(t)\right)^{2}
\right)\label{L_aux_S}
\end{align}
\end{subequations}
$\mathcal{L}_{B}$ and $\mathcal{L}_{C}$ are the same as
those defined in eq.\ref{eq:S} and respectively.
At this stage, it is convenient to perform a Keldysh rotation. The 
fermionic Grassmann fields are transformed as:
\begin{subequations}
\begin{align}
d^{1} = \frac{1}{\sqrt{2}}\left(d^{+} + d^{-}\right) &\,\,\,\,\,\,\,
\, \bar{d}^{1} = \frac{1}{\sqrt{2}}\left(\bar{d}^{+} - 
\bar{d}^{-}\right)\\
d^{2} = \frac{1}{\sqrt{2}}\left(d^{+} - d^{-}\right) &\,\,\,\,\,\,\,
\, \bar{d}^{2} = \frac{1}{\sqrt{2}}\left(\bar{d}^{+} 
+ \bar{d}^{-}\right)
\end{align}
\end{subequations}

while the bosonic auxiliary fields transform as:

\begin{align}
    M^{c} &= \frac{1}{2}\left(M^{+} + M^{-}\right)\\
    M^{q} &= \left(M^{+} - M^{-}\right)
\end{align}
where we have suppressed the time and other labels for notational 
brevity. A similar transformation for the $\phi$ field is given in 
eq.\ref{eq:kelrot}. The fermions can be integrated out and the 
resulting action in frequency space can be written as:
\begin{subequations}\label{S_diss}
\begin{align}
\tilde{S}\left[\phi,M\right] &= Tr\ln\left(i\check{G}^{-1}
(\omega,\omega^{\prime})\right)\nonumber\\
&+ \frac{1}{U}\int\frac{\mathrm{d}\omega}{2\pi}\sum_{i}
\left(\phi^{c}_{i}(\omega)\phi^{q}_{i}(-\omega) - M^{c}_{i}(\omega)
M^{q}_{i}(-\omega)\right)
\end{align}
where the components of $\check{G}^{-1}$ in the 2 $\times$ 2 Keldysh space are given by:
\begin{align}
\left[\left(\hat{G}^{-1}\right)^{R}\right]_{\substack{ij;\\
\alpha\alpha^{\prime}}} &= (\omega\delta_{ij} + t_{<ij>}
+i \hat{\Gamma}^{R}_{ij\alpha}(\omega))\delta_{\alpha\alpha^{\prime}}
\delta(\omega-\omega^{\prime})\nonumber\\
&- (\phi^{c}_{i}(\omega-\omega^{\prime})\delta_{\alpha\alpha^{\prime}}
- M^{c}_{i}(\omega-\omega^{\prime})\sigma^{z}_{\alpha\alpha^{\prime}})
\delta_{ij}\label{Ginv_diss:R}\\
\left[\left(\hat{G}^{-1}\right)^{12}\right]_{\substack{ij;\\
\alpha\alpha^{\prime}}} &= \frac{1}{2}(M^{q}_{i}(\omega
-\omega^{\prime})\sigma^{z}_{\alpha\alpha_{\prime}} 
- \phi^{q}_{i}(\omega-\omega^{\prime})\delta_{\alpha\alpha^{\prime}}
)\delta_{ij}\nonumber\\
&+ \hat{\Gamma}^{K}_{ij\alpha}(\omega)\delta(\omega-\omega^{\prime})
\delta_{\alpha\alpha^{\prime}}\label{Ginv_diss:12}\\
\left[\left(\hat{G}^{-1}\right)^{21}\right]_{\substack{ij;\\
\alpha\alpha^{\prime}}} &= \frac{1}{2}(M^{q}_{i}(\omega
-\omega^{\prime})\sigma^{z}_{\alpha\alpha^{\prime}} - \phi^{q}
(\omega-\omega^{\prime})\delta_{\alpha\alpha^{\prime}})
\delta_{ij}\label{Ginv_diss:21}\\
\left[\left(\hat{G}^{-1}\right)^{A}\right]_{\substack{ij;\\
\alpha\alpha^{\prime}}} &= (\omega\delta_{ij} + t_{<ij>}
+i \hat{\Gamma}^{A}_{ij\alpha}(\omega))\delta_{\alpha\alpha^{\prime}}
\delta(\omega-\omega^{\prime})\nonumber\\ 
&- (\phi^{c}_{i}(\omega-\omega^{\prime})\delta_{\alpha\alpha^{\prime}}
- M^{c}_{i}(\omega-\omega^{\prime})\sigma^{z}_{\alpha\alpha^{\prime}})
\delta_{ij}\label{Ginv_diss:A}
\end{align}
\end{subequations}
$\Gamma^{R,A,K}(\omega)$ are dissipation terms which enter the 
action as a result of integrating out the leads.\\

At this point one can, in principle, find the saddle point of this 
action with respect to the frequency dependent auxiliary fields. 
But for evaluating the saddle point equations one would have to invert 
the full frequency and site off-diagonal Green's function 
$\mathbb{G}^{-1}$. This is a very challenging task, and requires some 
physically motivated approximations in order to proceed. Here we 
invoke the static approximation, in which we drop the frequency 
dependence of the auxiliary fields, and thus restrict ourselves to the 
description of steady states only. The advantage we gain is that the 
Green's functions become diagonal in frequency, owing to the 
time-translation invariance of the NESS. This allows us to access much 
larger sizes, in order to establish detailed spatial dependence of 
various quantities of interest.

\section{Justification for collinear moments}\label{appendixB}

In this section we derive the form of the consistency conditions in 
the strong coupling limit. This allows us to cast the consistency 
conditions as approximate polynomial equations in terms of the 
auxiliary fields, with coefficients which are determined by tracing 
over the electrons.\\
\par
We can rewrite the $G^{-1}$ in eqn. \ref{G_SPA} as:
\begin{subequations}
\begin{align}
\check{G}^{-1}\left(\omega\right) = \check{\mathcal{G}}^{-1}\left(\omega\right) 
+ \check{\Sigma}
\end{align}
with,
\begin{align}
&\check{\mathcal{G}}^{-1}\left(\omega\right) =\nonumber\\
&\begin{bmatrix}(\omega-\hat{\Lambda}^{c}_{i})\delta_{ij}
+\hat{\Gamma}^{R}_{ij\alpha}\left(\omega\right)\delta_{\alpha\alpha^{\prime}}
&\hat{\Gamma}^{K}_{ij\alpha}\left(\omega\right)
\delta_{\alpha\alpha^{\prime}} \\
0 & (\omega-\hat{\Lambda}^{c}_{i})\delta_{ij}+\hat{\Gamma}^{A}_{ij}
\left(\omega\right)\delta_{\alpha\alpha^{\prime}}
\end{bmatrix}\\
\mbox{and,}\nonumber\\
\check{\Sigma} &= \begin{bmatrix}t_{<ij>} & \hat{\Lambda}^{q}_{i}\delta_{ij}\\
\hat{\Lambda}^{q}_{i}\delta_{ij} & t_{<ij>}
\end{bmatrix} = \hat{T}\mathcal{I}^{K} + \hat{\Lambda}^{q}\sigma^{K}_{1}
\end{align}
\end{subequations}
With this, we can write the SPA action given in eqn.\ref{S_SPA} in 
the following way:
\begin{subequations}
\begin{align}
S^{SPA}\left[\phi,\bm{M}\right] &=-i Trln
\left[i\check{\mathcal{G}}^{-1}\right]-i Trln\left[\mathcal{I}
+\check{\mathcal{G}}\check{\Sigma}\right]+S^{\prime}
\end{align}
The first term vanishes due to the causal structure of the theory. We 
expand the second term in powers of $\Sigma$.
\begin{align}
S^{SPA} &= S^{(1)} + S^{(2)} + S^{(3)} + O(\Sigma^{4}) + S^{\prime}
\end{align}
with 
\begin{align}
S^{(n+1)} &= i Tr\left(\left(\hat{\mathcal{G}}\hat{T}\right)^{n}
\hat{\mathcal{G}}\hat{\Lambda}^{q}\hat{\sigma}^{K}_{1}\right) + \dots
\end{align}
\end{subequations}

The even powers of this expansion, along with the terms denoted by 
ellipsis vanish at the classical saddle point. We truncate the expansion
to $O(t^{2})$ and evaluate the classical saddle point. In this limit 
the saddle point equations are:
\begin{subequations}\label{pert:main}
\begin{align}
\frac{\delta S^{SPA}}{\delta\phi^{q}_{i}}&\at[\bigg]{\{q=0\}} = 
\frac{i}{2}Tr\left(\hat{\mathcal{G}}^{K}_{i}\right)
+\frac{2}{U}\phi^{c}_{i}-\frac{it^{2}}{2}\sum\limits_{z\in NN}
\nonumber\\
&Tr\Biggl(\hat{\mathcal{G}}^{K}_{i}\left(\hat{\mathcal{G}}^{R}_{i}
\hat{\mathcal{G}}^{R}_{i+z}+\hat{\mathcal{G}}^{A}_{i+z}\hat{\mathcal{G}}^{A}_i\right)
+ \hat{\mathcal{G}}^{K}_{i+z}\hat{\mathcal{G}}^{A}_{i}\hat{\mathcal{G}}^{R}_{i}\Biggr)
\nonumber\\
&= 0\label{pert:a}\\
\frac{\delta S^{SPA}}{\delta \bm{M}^{q}_{i}}&\at[\bigg]{\{q=0\}} = 
-\frac{i}{2}Tr\left(\hat{\mathcal{G}}^{K}_{i}\bm{\sigma}\right)
- \frac{1}{U}\bm{M}^{c}_{i}+\frac{it^{2}}{2}\sum\limits_{z\in NN}
\nonumber\\
&Tr\Biggl(\hat{\mathcal{G}}^{K}_{i}\left(\bm{\sigma}\hat{\mathcal{G}}^{R}_{i}
\hat{\mathcal{G}}^{R}_{i+z}+\hat{\mathcal{G}}^{A}_{i+z}\hat{\mathcal{G}}^{A}_{i}
\bm{\sigma}\right) + \hat{\mathcal{G}}^{K}_{i+z}\hat{\mathcal{G}}^{A}_{i}
\bm{\sigma}\hat{\mathcal{G}}^{R}_{i}\Biggr)\nonumber\\
&= 0\label{pert:b}
\end{align}
\end{subequations}
where $\hat{\mathcal{G}}^{R,A,K}_{i}$ are the atomic Green's functions which 
can be expressed in terms of $\phi^{c}$ and $\bm{M}^{c}$. `NN' denotes
the nearest neighbours. The trace involves a matrix trace over the 
local spin components as well as an integral over frequency, which can
be performed analytically. Hence, these constitute a complete set of 
equations which determine the auxiliary field profile over the entire 
system, as long as we stay in the regime $t^{2} < U|M^{c}_{i}|$. In 
principle, they can be solved to obtain the $\phi$, as well as the 
$\bm{M}$ profile throughout the system.\\
\par
For our purposes, it is sufficient to focus on eqn. \ref{pert:b}. The 
magnitude of the local moment is obtained by equating the sum of first 
two terms to zero. Demanding that the third term should vanish 
determines the relative phase between neighbouring spins. After taking
the trace over the spin indices and integrating over frequency, it 
simplifies to:
\begin{align}
\left(\left(\bm{M}_{i}\cdot\bm{M}_{j}\right)\bm{M}_{i}
-|\bm{M}_{i}|^2\bm{M}_{j}\right)f\left(\phi_{i},|M_{i}|,\phi_{j},|M_{j}|,U\right) = 0
\end{align}
where $i$ and $j$ are nearest neighbour sites and 
$f\left(\phi_{i},|M_{i}|,\phi_{j},|M_{j}|,U\right)$ 
is independent of relative orientation of the moments. This can be 
satisfied for any value of the $\{\phi,|M|\}$, provided 
$\left(\left(\bm{M}_{i}\cdot\bm{M}_{j}\right)\bm{M}_{i}-|\bm{M}_{i}|^2\bm{M}_{j}
\right) = 0$.
We can easily verify that both FM, as well as a N\'eel AFM state 
satisfies this condition. For the square lattice, we know that the 
N\'eel state is lower in energy at equilibrium, and is energetically 
well separated from the FM state. This serves as the justification for 
neglecting the $M^{c}_{x}$, $M^{c}_{y}$ components of the consistency 
equation \ref{MF:M} in numerical implementation of the mean field 
scheme.

\section{Perturbative corrections to DOS}
\label{appendixC}
Assume the bare Hamiltonian to be translation invariant.
\begin{equation}
H_{0} = \sum\limits_{\substack{{\bf k}\in\frac{1}{2}BZ\\\sigma\in\pm}}
\begin{pmatrix}c^{\dagger}_{{\bf k}\sigma} &
 c^{\dagger}_{{\bf k+Q}\sigma}\end{pmatrix}
\begin{bmatrix}\epsilon_{\bf k} & \sigma M_0 \\
\sigma M_0 & -\epsilon_{\bf k} \end{bmatrix}
\begin{pmatrix}c_{{\bf k}\sigma} \\ c_{{\bf k+Q}\sigma}\end{pmatrix}
\end{equation}
where ${\bf Q} = (\pi,\pi)$. $M_{0}$ is the moment size which can be
obtained by solving the gap equation. This can be diagonalised using
the transformation:
\begin{equation}
\label{eq:boguliubov}
\begin{pmatrix}\gamma^{A}_{{\bf k},\sigma}\\\gamma^{B}_{{\bf k},\sigma}\end{pmatrix}
= \begin{pmatrix}\cos\theta_{\bf k} & \sin\theta_{\bf k}\\\sin\theta_{\bf k}
& -\cos\theta_{\bf k}\end{pmatrix}
\begin{pmatrix}c_{{\bf k},\sigma} \\ c_{{\bf k+Q},\sigma}\end{pmatrix}
\end{equation}
with $\theta_{\bf k} = \tan^{-1}\left(\frac{E_{\bf k}-\epsilon_{\bf k}}{M_{0}}\right)$
and $E_{\bf k} = \sqrt{\epsilon_{\bf k}^{2} + M_{0}^{2}}$.
This maps the problem into a 2-band system separated by a gap.
The retarded Green's functions for the upper and the lower bands are given by:
\begin{equation}
\begin{gathered}
  \left[\mathcal{G}^{R}(\omega)\right]^{AA}_{\bf k} =
  \feynmandiagram[horizontal=a to b]
{a -- [fermion,edge label=\({\bf k}\)] b}; = \frac{1}{\omega-E_{\bf k}+i\eta}
\end{gathered}
\end{equation}
\begin{equation}
\begin{gathered}
\left[\mathcal{G}^{R}(\omega)\right]^{BB}_{\bf k} =
\feynmandiagram[horizontal=a to b]
{a -- [charged scalar,edge label=\({\bf k}\)] b};= \frac{1}{\omega+E_{\bf k}+i\eta}
\end{gathered}
\end{equation}
We shall suppress the label $`R'$, henceforth in this section, as all the Green's
functions and the self-energies shall pertain to the Retarded component.
The bias acts perturbatively on the system via the charge and spin deviation
fields, given in eq.\ref{eq:auxdev}.
Implementing a discrete fourier transform on the a lattice:
\begin{align}
\delta\phi(k_{x,}k_{y}) &= \sum\limits_{x=-L_{x}/2}^{L_{x}/2}
\sum\limits_{y=-L_{y}/2}^{L_{y}/2}e^{i(k_{x}x+k_{y}y)}\delta\phi(x)\nonumber\\
&= \frac{\delta_{k_{y},0}}{L_{x}}\int\limits_{-L_{x}/2}^{L_{x}/2}\mathrm{d}x
\,e^{ik_{x}x}\delta\phi(x)\nonumber\\
&= \frac{2A}{L_{x}}e^{-\frac{L_{x}}{2\xi}}\delta_{k_{y},0}
\Biggl[\frac{sinh\left((k_{0}-ik_{x})\frac{L_{x}}{2}\right)}{(k_{0}-ik_{x})}\nonumber\\
&-\frac{sinh\left((k_{0}+ik_{x})\frac{L_{x}}{2}\right)}{(k_{0}+ik_{x})}\Biggr]
\end{align}
for $a_{x}/L_{x} << 1$, where $a_{x}$ is the lattice spacing in the x-direction.
Similarly,
\begin{align}
&\delta M(k_{x},k_{y}) = -\frac{A}{L_{x}}e^{-\frac{L_{x}}{2\xi}}\Biggl[\delta_{k_{y}^{+},0}
\Biggl(\frac{sinh\left((k_{0}-ik_{x}^{+})\frac{L_{x}}{2}\right)}{(k_{0}-ik_{x}^{+})}\nonumber\\
&+\frac{sinh\left((k_{0}+ik_{x}^{+})\frac{L_{x}}{2}\right)}{(k_{0}+ik_{x}^{+})}\Biggr)
+ \Bigl((k_{x}^{+},k_{y}^{+})\rightarrow (k_{x}^{-},k_{y}^{-})\Biggr)\Biggr]
\end{align}
where $k_{x,y}^{\pm} = k_{x,y}\pm\pi$.
The perturbation can be written in terms of the deviation fields as:
\begin{align}
\label{eq:H_pert}
H_{pert}&\nonumber\\
 =& \sum\limits_{\substack{{\bf k},{\bf q}\in\frac{1}{2}BZ\\\sigma\in\pm}}
    \begin{pmatrix}c^{\dagger}_{{\bf k}\sigma} &
      c^{\dagger}_{{\bf k+Q}\sigma}\end{pmatrix}
    \begin{bmatrix}\delta\phi_{\bf k-q} & \sigma\delta M_{\bf k-q} \\
      \sigma\delta M_{\bf k-q} & -\delta\phi_{\bf k-q} \end{bmatrix}
    \begin{pmatrix}c_{{\bf q}\sigma} \\ c_{{\bf q+Q}\sigma}\end{pmatrix}\nonumber\\
  +& H.c. \\
 =& \sum\limits_{\substack{{\bf k},{\bf q}\in\frac{1}{2}BZ\\\sigma\in\pm}}
    \begin{pmatrix}\left(\gamma^{A}_{{\bf k}\sigma}\right)^{\dagger} &
      \left(\gamma^{B}_{{\bf k}\sigma}\right)^{\dagger}\end{pmatrix}
    \begin{bmatrix}f({\bf k},{\bf q}) & g({\bf k},{\bf q}) \\
      g({\bf k},{\bf q}) & -f({\bf k},{\bf q}) \end{bmatrix}
 \begin{pmatrix}\gamma_{{\bf q}\sigma} \\ \gamma_{{\bf q}\sigma}\end{pmatrix}\nonumber\\
  +& H.c. \\
  \mbox{where,}\nonumber\\
  f({\bf k},{\bf q}) &= cos(\theta_{\bf k}+\theta_{\bf q})\delta\phi_{\bf k-q} +
  \sigma sin(\theta_{\bf k}+\theta_{\bf q})\delta M_{\bf k-q}\\
  g({\bf k},{\bf q}) &= sin(\theta_{\bf k}+\theta_{\bf q})\delta\phi_{\bf k-q} -
  \sigma cos(\theta_{\bf k}+\theta_{\bf q})\delta M_{\bf k-q}
  \end{align}                      
Self-energy: The first order self energy vanishes, since there is no
  ${\bf k} = 0$ component to the deviation fields. The second order self
  energy correction to the upper and lower band propagators is given by:
\begin{align}
    \Sigma^{AA}_{{\bf k}\sigma}(\omega) &=
  \feynmandiagram[horizontal=a to b,baseline=(a.base)]
  {a -- [fermion, edge label'=\({\bf q}\)] b,
  a -- [charged boson, half left, edge label=\({\bf k-q}\)] b};\,\,+\,\,
\feynmandiagram[horizontal=a to b,baseline=(a.base)]
{a -- [charged scalar,edge label'=\({\bf q}\)] b,
      a -- [gluon, half left, edge label=\({\bf k-q}\)] b};\nonumber\\
    &= \sum\limits_{{\bf q}\in\frac{1}{2}BZ}\left(
      \frac{|f({\bf k},{\bf q})|^{2}}{\omega-E_{\bf q}+i\eta} +
    \frac{|g({\bf k},{\bf q})|^{2}}{\omega+E_{\bf q}+i\eta}\right)
\end{align}

\begin{align}
 \Sigma^{BB}_{{\bf k}\sigma}(\omega) &=
 \feynmandiagram[horizontal=a to b,baseline=(a.base)]
 {a -- [charged scalar, edge label'=\({\bf q}\)] b,
      a -- [charged boson, half left, edge label=\({\bf k-q}\)] b};
 \,\,+\,\,
 \feynmandiagram[horizontal=a to b,baseline=(a.base)]
{a -- [fermion,edge label'=\({\bf q}\)] b,
      a -- [gluon, half left, edge label=\({\bf k-q}\)] b};\nonumber\\
&= \sum\limits_{{\bf q}\in\frac{1}{2}BZ}\left(
      \frac{|f({\bf k},{\bf q})|^{2}}{\omega+E_{\bf q}+i\eta} +
    \frac{|g({\bf k},{\bf q})|^{2}}{\omega-E_{\bf q}+i\eta}\right)
\end{align}
The corrected propagators are given by:
  \begin{align}
    G^{AA}_{{\bf k}\sigma}(\omega) = \left(\left[\mathcal{G}^{AA}_{{\bf k}\sigma}
    (\omega)\right]^{-1}
    -\Sigma^{AA}_{{\bf k}\sigma}(\omega)\right)^{-1}\\
    G^{BB}_{{\bf k}\sigma}(\omega) = \left(\left[\mathcal{G}^{BB}_{{\bf k}\sigma}
    (\omega)\right]^{-1}
    -\Sigma^{BB}_{{\bf k}\sigma}(\omega)\right)^{-1}
  \end{align}
  From this the perturbatively corrected DOS can be obtained by:
  \begin{equation}
    \label{eq:corr_dos}
    \mathcal{A}_{pert}(\omega) = -\frac{1}{\pi}\sum\limits_{{\bf k}\sigma}
    Im\left[G^{AA}_{{\bf k}\sigma}(\omega) + G^{BB}_{{\bf k}\sigma}(\omega)\right]
  \end{equation}

  
\bibliographystyle{apsrev4-1}
%
\end{document}